\documentclass[twocolumn,nofootinbib,showpacs,preprintnumbers,superscriptaddress]{revtex4}
\usepackage[dvipdfmx]{graphicx}
\usepackage{amssymb,amsmath,amsfonts,amsbsy}
\usepackage{color}
\usepackage{enumerate} 
\usepackage{ulem}
\usepackage{multirow}

\begin{document}

\title{Probing primordial non-Gaussianity with 21~cm fluctuations from minihalos}

\author{Toyokazu Sekiguchi}
\email{sekiguti@resceu.s.u-tokyo.ac.jp}
 \affiliation{Research Center for the Early Universe (RESCEU), Graduate School of Science, the University of Tokyo, Tokyo 113-0033, Japan}

\author{Tomo Takahashi}
\email{tomot@cc.saga-u.ac.jp}
\affiliation{Department of Physics, Saga University, Saga 840-8502, Japan}

\author{Hiroyuki Tashiro}
\email{hiroyuki.tashiro@nagoya-u.jp}
 \affiliation{Department of Physics, Graduate School of Science, Nagoya University, Aichi 464-8602, Japan}
 
\author{Shuichiro Yokoyama}
\email{shuichiro@rikkyo.ac.jp}
 \affiliation{Department of Physics, Rikkyo University, 3-34-1 Nishi-Ikebukuro, Toshima, Tokyo 171-8501, Japan}
 \affiliation{Kavli Institute for the Physics and Mathematics of the Universe
(Kavli IPMU, WPI), Todai Institutes for Advanced Study, the University of
Tokyo, Kashiwa 277-8583, Japan}

\begin{abstract}

We investigate future constraints on primordial local-type non-Gaussianity from 21~cm angular power spectrum from minihalos.
{We particularly focus on the trispectrum of primordial curvature perturbations which are characterized by the non-linearity 
parameters $\tau_{\rm NL}$ and $g_{\rm NL}$.}
We show that  future measurements of minihalo 21~cm angular power spectrum can probe these non-linearity parameters 
with an unprecedented precision of $\tau_{\rm NL} \sim 30$ and $g_{\rm NL} \sim 2 \times 10^3$ for Square Kilometre Array~(SKA) and 
$\tau_{\rm NL} \sim 0.6$ and $g_{\rm NL} \sim 8 \times 10^2$ for Fast Fourier Transform Telescope~(FFTT). 
These levels of sensitivity would give significant implications for models of the inflationary Universe and 
the origin of cosmic density fluctuations.

\end{abstract}


\preprint{RUP-18-19 RESCEU-9/18}

\maketitle


{\it Introduction}.---\quad 
Primordial non-Gaussianity is one of the most important quantities to probe the inflationary Universe and 
the origin of density fluctuations.
Its existence inevitably indicates an inflationary model beyond the standard single-field inflation  such as the existence of multiple fields, 
noncanonical kinetic term,  deviations from the initial Bunch-Davies vacuum and so on~(see, {\it e.g.}, Ref~\cite{Bartolo:2004if,Chen:2010xka,Wands:2010af} for reviews and references therein).
Non-Gaussianities can be characterized by bispectrum $B$ and trispectrum $T$, which
are defined by connected part of three point and four point functions  of the primordial 
curvature perturbation $\Phi (\vec{k})$:
\begin{align}
& \left\langle \Phi ( \vec{k}_1) \Phi (\vec{k}_2 ) \Phi (\vec{k}_3) \right\rangle = (2\pi)^3 B(\vec k_1, \vec k_2, \vec k_3) \delta (\vec{k}_1 + \vec{k}_2 + \vec{k}_3 ), \\ 
& \left\langle \Phi ( \vec{k}_1) \Phi (\vec{k}_2 ) \Phi (\vec{k}_3) \Phi (\vec{k}_4) \right\rangle_{\rm conn} \notag \\
&\qquad\qquad = (2\pi)^3 T( \vec k_1, \vec k_2, \vec k_3, \vec k_4) \delta (\vec{k}_1 + \vec{k}_2 + \vec{k}_3 +\vec{k}_4).
\end{align}
For the case of the so-called local-type non-Gaussianity,  we can expand $\Phi$ as~\cite{Komatsu:2001rj}
\begin{equation}
\Phi(\vec x)=\Phi_{\rm G}(\vec x)+f_{\rm NL}(\Phi_{\rm G}(\vec x)^2-\langle\Phi_{\rm G}\rangle^2) 
+g_{\rm NL}\Phi_{\rm G}(\vec x)^3,
\label{eq:local}
\end{equation}
and the bispectrum and the trispectrum are given by
\begin{align}
&B(\vec k_1,\vec k_2,\vec k_3)=2f_{\rm NL}[P_\Phi(k_1)P_\Phi(k_2)+\mbox{(2 perm)}], \label{eq:bispec}
\end{align}
\begin{align}
&T(\vec k_1,\vec k_2,\vec k_3,\vec k_4)\notag \\
&\quad\quad=6g_{\rm NL}[P_\Phi(k_1)P_\Phi (k_2) P_\Phi (k_3)+\mbox{(3 perm)}] \notag \\ 
&\quad\quad+\frac{25}9\tau_{\rm NL}[P_\Phi (k_1)P_\Phi(k_2) P_\Phi(k_{13})+\mbox{(11 perm)}], \label{eq:trispec}
\end{align}
where $P_\Phi(k)$ is the power spectrum of Gaussian part of $\Phi$ and
$k_{13} \equiv \left|\vec k_1+\vec k_3\right|$.
The parameters $f_{\rm NL}$ and $g_{\rm NL}$ correspond to 
the amplitude of bispectrum $B$ and trispectrum $T$ normalized by $P_\Phi^2$ and
 $P_\Phi^3$, respectively. Besides, $\tau_{\rm NL}$ is another parameter which characterizes the size of the trispectrum for a different configuration of wave numbers.

To date, cosmic microwave background~(CMB) observations give the most severe constraint on non-Gaussianities.
The current best constraints on $f_{\rm NL}$ and $g_{\rm NL}$ are given by the Planck 2015 result as 
$f_{\rm NL}=0.8\pm5.0$ and $g_{\rm NL}=(9.0\pm7.7)\times10^4$ at 1$\sigma$~\cite{Ade:2015ava},
whilst the one for $\tau_{\rm NL}$ is given as $\tau_{\rm NL}\le 2800$ at 95\%~C.L. by the Planck 2013 result \cite{Ade:2013ydc}.
Although the constraint on $f_{\rm NL}$ is relatively severe, the ones for the trispectrum, $\tau_{\rm NL}$ and $g_{\rm NL}$, are 
rather weak and do not give meaningful information for models of the inflationary Universe. 
However, the trispectrum is potentially very important to differentiate inflationary models since they provide a consistency check
of the models.
In fact, when the source of primordial fluctuations originates from a single field, 
$\tau_{\rm NL}$ in Eq.~\eqref{eq:trispec} is related to $f_{\rm NL}$ by $\tau_{\rm NL}=(36/25)f_{\rm NL}^2$~\cite{Boubekeur:2005fj}.
However, on the other hand, when primordial fluctuations are generated  from multiple fields, the above relation becomes  an inequality $\tau_{\rm NL}\ge (36/25) f_{\rm NL}^2$,
which is the so-called Suyama-Yamaguchi inequality \cite{Suyama:2007bg}, and 
this inequality has been shown to be valid under a quite general assumption \cite{Smith:2011if,Assassi:2012zq} which are  
satisfied in almost all models of primordial fluctuations suggested so far. 
In any case, the deviation from the relation $\tau_{\rm NL}=(36/25)f_{\rm NL}^2$ would give significant implications for the inflationary Universe
and hence checking this relation is very important, where the information of the trispectrum is essential.  
Furthermore, even if the amplitude of the bispectrum $f_{\rm NL}$ is small, the trispectrum can be large in some models \cite{Suyama:2010uj,Suyama:2013rol}.
In this regard, it would be worth investigating to what extent we can probe the trispectrum in  future observations. 
We in this paper study expected sensitivities for the non-linearity parameters, particularly focusing on $\tau_{\rm NL}$ and $g_{\rm NL}$ from 
future 21~cm line fluctuations from minihalos by adopting Fisher matrix analysis for its angular power spectrum.

Redshifted 21~cm line emission/absorption is the unique probe of cosmic neutral hydrogen in the dark ages.
By observing its fluctuations, we can extract cosmological information from unexplored redshifts
with an unprecedented volume~\cite{Furlanetto:2006jb}. Following the recent detection reported by 
EDGES~\cite{Bowman}, upcoming Square Kilometre Array (SKA)~\cite{SKA}
will be measuring the 21~cm fluctuations which is expected to enhance our understanding of 
the early Universe~\cite{Blake:2004pb}.
The feasibility study to constrain the primordial non-Gaussianity
parameters through 21~cm observations has been performed with considering the
various aspects of 21~cm signatures of the primordial non-Gaussianity~\cite{Cooray:2006km,Joudaki:2011sv,Tashiro:2011br,Tashiro:2012wr,Chongchitnan:2012we}.
In this paper we particularly focus on the 21~cm line fluctuations from the 
so-called minihalos~\cite{Iliev:2002gj,Furlanetto:2002ng,Sekiguchi:2013lma,Sekiguchi:2014wfa}, 
which are halos so small that its virial temperature is not high enough to activate star formation inside.
While much attention has been paid to the 21~cm fluctuations from smoothed intergalactic medium, minihalos
are expected to contribute predominantly at low redshifts near the completion of the cosmic reionization.
With its abundance being sensitive to density fluctuations at sub-Mpc scales, 21~cm line fluctuations 
from minihalos enable us to probe primordial perturbations at a wide range of scales~\cite{Sekiguchi:2017cdy}.

\bigskip
{\it 21~cm angular power spectrum from minihalos}.---\quad
Given a  line of sight $\hat n$ and a redshift $z$,  fluctuations in the differential brightness 
temperature $\delta  \Delta T_b(\hat n, z)$ from minihalos is given by~\cite{Iliev:2002gj,Sekiguchi:2013lma,Sekiguchi:2014wfa}
\begin{eqnarray}
\delta  \Delta T_b(\hat n, z)&=&\overline{\Delta T_b}(z) 
\delta_h(\vec x=r(z)\hat n, z), \label{eq:deltadiffT}
\end{eqnarray}
where $\overline{\Delta T_b}$ is the mean differential brightness temperature and $\delta_h$ is the fractional overdensity 
in the minihalo number density. 
Note that, in Eq.~\eqref{eq:deltadiffT}, we omitted the redshift-space distortion, which will be incorporated shortly later.
On large scales, $\delta_h$ is linearly related to the matter overdensity $\delta$ in the Fourier space as
\begin{eqnarray}
\delta_h(\vec k, z)&=&\beta(k, z)\delta(\vec k, z), \label{eq:delh}
\end{eqnarray}
where $\beta$ is the effective bias of minihalos with respect to the underlying matter density fluctuations $\delta$. 
The bias $\beta$ is given by~\cite{Iliev:2002gj}
\begin{equation}
\beta(k,z)\equiv \frac{\displaystyle\int^{M_{\rm max}}_{M_{\rm min}}  dM \frac{dn}{dM} \mathcal F(z,M) b(k,M,z)}
{\displaystyle\int^{M_{\rm max}}_{M_{\rm min}}  dM \frac{dn}{dM} \mathcal F(z,M)},
\label{eq:ave-bias}
\end{equation}
where $b(k,M,z)$ is the bias of minihalos with mass $M$, $\mathcal F(z,M)$ is a flux from a single minihalo and $dn/dM$ is the mass function of minihalos.
The halo matter power spectrum can be given by
\begin{equation}
P_{hh}(k; z,z')=\beta(k,z)\beta(k,z')D(z)D(z')P_{\delta\delta}(k),
\label{eq:hpz0}
\end{equation}
with $D(z)$ being the growth factor at $z$ normalized to unity at $z=0$, where the the matter power spectrum 
$P_{\delta\delta}(k)$ is measured.

In the presence of the local-type non-Gaussianity, the local number density of halos is modulated 
by the long-wavelength fluctuations.
This leads to the scale-dependence in the halo bias at very large scale. The deviation in 
$b(k,M,z)$ (in $\beta (k,z)$) from the Gaussian case is 
given by \cite{Dalal:2007cu,Matarrese:2008nc,Slosar:2008hx,Smith:2011ub,Gong:2011gx,Yokoyama:2012az}
\begin{eqnarray}
\Delta b(k,M,z)&\approx&\frac{
\left\{b_f(M,z)f_{\rm NL}+b_g(M,z)g_{\rm NL}\right\}}{\alpha(k,z)}, \\
\alpha(k,z)&\equiv&
\frac{2k^2T(k)D(z)}{3\Omega_mH_0^2},
\end{eqnarray}
where $T(k)$ is the transfer function, 
$\Omega_m$ is the density parameter for total matter and $H_0$ is the Hubble constant.
We use the following expressions for $b_f$ and $b_g$~\cite{Smith:2011ub}:
\begin{eqnarray}
b_f (M,z)&=&2[b_0(M,z)-1]\delta_{\rm cr}, \\
b_g (M,z)&=&\hat \kappa_3\left[-1+\frac32(\nu-1)^2+\frac12(\nu-1)^3\right]\notag \\
&&\quad+\frac{d\hat \kappa_3}{d\log \sigma}\left(\frac{\nu-\nu^{-1}}2\right), 
\end{eqnarray}
where $\delta_{\rm cr}\simeq1.67$, $\nu=\delta_{\rm cr}/\sigma$, 
and we denote the root mean square of matter density fluctuations smoothed over a
top-hat volume enclosing mass $M$ by $\sigma$.
Here, $b_0(M,z)$ is the Gaussian linear bias
and $\hat \kappa_3$ is the (reduced) third order cumulant defined by $f_{\rm NL}\hat \kappa_3\equiv\langle\delta^3\rangle/\sigma^3$:
\begin{eqnarray}
\hat \kappa_3&=&\frac6{\sigma^3}\int \frac{d^3k_1}{(2\pi)^3}
\frac{d^3k_2}{(2\pi)^3}
W_M(k_1)W_M(k_2)W_M(k_{12}) \notag \\
&&\times\frac{P_{\delta\delta}(k_1)P_{\delta\delta}(k_2) \alpha(k_{12})}
{\alpha(k_1)\alpha(k_2)},
\end{eqnarray}
with $W_M(k)$ being the window function corresponding to a mass $M$.
For the purpose of demonstration, in this paper we adopt the Sheth-Tormen mass function~\cite{Sheth:1999mn}
in computing the minihalo abundance and its Gaussian linear bias.
Note that here we have neglected effects of the primordial non-Gaussianity on the 
mean minihalo abundance, because the effects should not be significant~\cite{Yokoyama:2011sy}
for the level non-Gaussianity we suppose in this paper.

By taking into account the redshift space distortion at linear level (i.e.~the Kaiser effect~\cite{Kaiser:1987qv}), 
the fractional overdensity of minihalo abundance in redshift space (denoted with $\delta_h^s$)
is  given by
\begin{equation}
\delta^{s}_h(\vec k,z)=\delta_h(\vec k,z)+f(z)\mu^2\delta(\vec k,z),
\end{equation}
where $f(z)=d\ln D(z)/d\ln a$ and $\mu=\hat k\cdot \hat n$ is the cosine between $\vec k$ and the line of sight $\hat n$.

So far, we have been adopting the exact form of Eq.~\eqref{eq:local} for $\Phi$
and this results in 
a term proportional to $f_{\rm NL}^2$ in the expression of Eq.~\eqref{eq:hpz0}.
In general, multiple sources contribute to non-Gaussianity and in such a case
one needs to replace $f_{\rm NL}^2$ with $({25}/36)\tau_{\rm NL}$.
Finally, we obtain the following expression for the minihalo power spectrum in the redshift space as 
\begin{widetext}
\begin{eqnarray}
P^s_{hh}(k; z,z')&\approx&
\Big[\{\beta_0(z)+f(z)\mu^2\}\{\beta_0(z')+f(z')\mu^2\}
+\frac{\Delta\beta(z)}{\alpha(k,z)}\{\beta_0(z')+f(z')\mu^2\}\notag \\
&&\qquad\qquad 
+\{\beta_0(z)+f(z)\mu^2\}\frac{\Delta\beta(z')}{\alpha(k,z')}
+\frac{36}{25}\tau_{\rm NL}\frac{\beta_f(z)\beta_f(z')}{\alpha(k,z)\alpha(k,z')} \Big]D(z)D(z')P_{\delta\delta}(k), 
\label{eq:hpz} 
\end{eqnarray}
\end{widetext}
\begin{eqnarray}
\Delta \beta (z) &=& \beta_f (z) f_{\rm NL} + \beta_g (z) g_{\rm NL},
\end{eqnarray}
where $\beta_0$, $\beta_f$ and $\beta_g$ are obtained by replacing 
$b(k,M,z)$ in Eq.~\eqref{eq:ave-bias} with $b_0$, $b_f$ and $b_g$, respectively.
Note that here we have omitted terms proportional to $f_{\rm NL}g_{\rm NL}$ or $g_{\rm NL}^2$ 
since they would give minor contributions.

In the same manner as in Ref.~\cite{Sekiguchi:2017cdy}, 
we define the tomographic angular power spectrum of the 
21cm line fluctuations from minihalos, $C_l^{\rm (21cm)}(z,z')$ by 
\begin{eqnarray}
\langle a_{lm}
(z) a_{l'm'}
(z')\rangle&=&
C_l^{\rm (21cm)}(z,z')\delta_{ll'}\delta_{mm'}, 
\label{eq:defCl} \\
a_{lm}
(z)&\equiv&\int d\hat n\,\delta \Delta T_b(\hat n, z) Y^*_{lm}(\hat n).
\end{eqnarray}

\bigskip
{\it Forecasted constraints}.---\quad
We perform the Fisher matrix analysis to obtain forecasted constraints
on $f_{\rm NL}, \tau_{\rm NL}$ and $g_{\rm NL}$.
Details of the computation of the Fisher matrix of 21~cm line and CMB angular power spectra are provided in our previous paper~\cite{Sekiguchi:2017cdy}. 
Specifications of surveys adopted in this paper are summarized in 
Tables~\ref{tab:cmb} 
\cite{Planck:2006aa,core} and \ref{tab:21cm} \cite{SKA,Tegmark:2008au}.
In our baseline analysis,
the maximum and minimum redshifts where minihalo can be observed are set to $z_{\rm max}=20$ and $z_{\rm min}=6$, respectively.
Since $z_{\rm min}$ can sizeably affect the forecasted constraints, we also examine the dependence of our results on $z_{\rm min}$.
In addition to the angular power spectra of CMB and 21~cm line, we also include
the CMB temperature bispectrum and trispectrum. We compute the CMB Fisher matrix of the non-linearity parameters based on
\cite{Komatsu:2001rj,Kogo:2006kh,Sekiguchi:2013hza}, neglecting the correlation 
between $g_{\rm NL}$ and $\tau_{\rm NL}$ for simplicity.

\begin{figure*}[tbp]
\centering 
   \includegraphics[width=.9\textwidth]{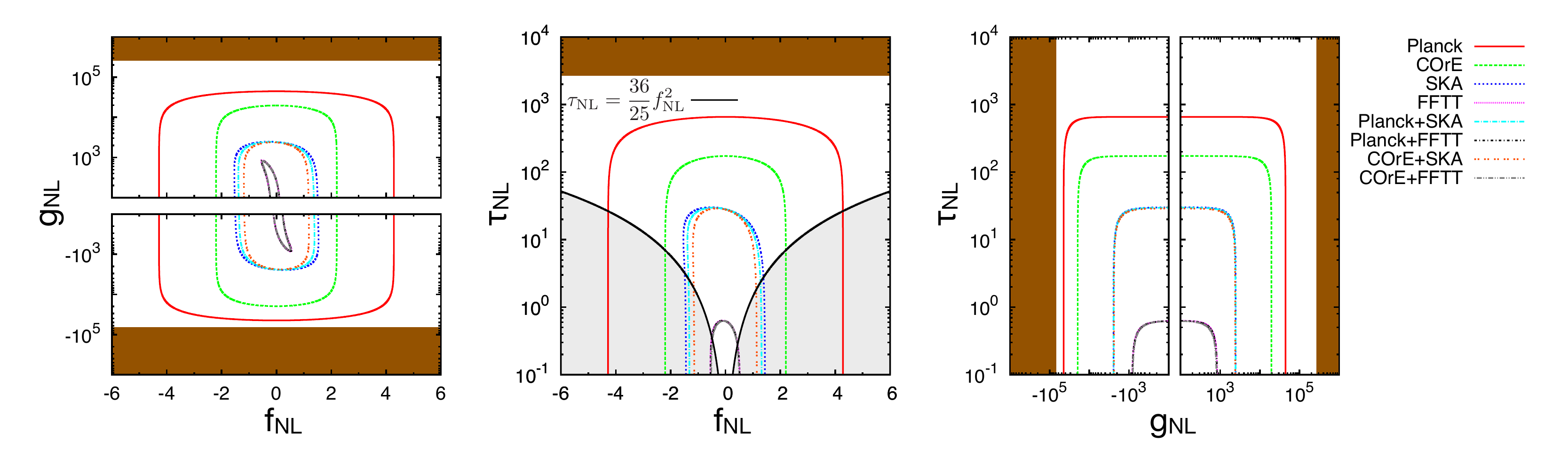}
\caption{\label{fig:fnlgnl} Expected 2-dim constraints on the nonlinearity parameters assuming $z_{\rm min}=6$.
Shaded region is excluded by current observations (see text for details).
}
\end{figure*}

The expected 1$\sigma$ errors based on the Fisher matrix analysis
are summarized in Table~\ref{tab:constraints}.
In Fig.~\ref{fig:fnlgnl}, projected constraints are shown on the $f_{\rm NL}$--$g_{\rm NL}$, $f_{\rm NL}$--$\tau_{\rm NL}$ and 
$g_{\rm NL}$--$\tau_{\rm NL}$ planes. 
For reference, the current 2$\sigma$ constraints on $\tau_{\rm NL}$ and $g_{\rm NL}$ are also shown by shade for the excluded parameter space. 
In the $f_{\rm NL}$--$\tau_{\rm NL}$ plane, the line of $\tau_{\rm NL} = (36/25) f_{\rm NL}^2$
and the region where the Suyama-Yamaguchi inequality does not hold are also shown.
As seen from the figure, future observations of 21~cm angular power spectrum can much improve constraints on the non-linearity parameters,
by a few orders of magnitude compared to the current ones.
Even compared with future CMB observations, the sensitivity is better already at the level of SKA. With the specification of FFTT, 
we can obtain unprecedented sensitivities particularly for $\tau_{\rm NL}$ and $g_{\rm NL}$.

\begin{figure}[tbp]
\centering 
   \includegraphics[width=.45\textwidth]{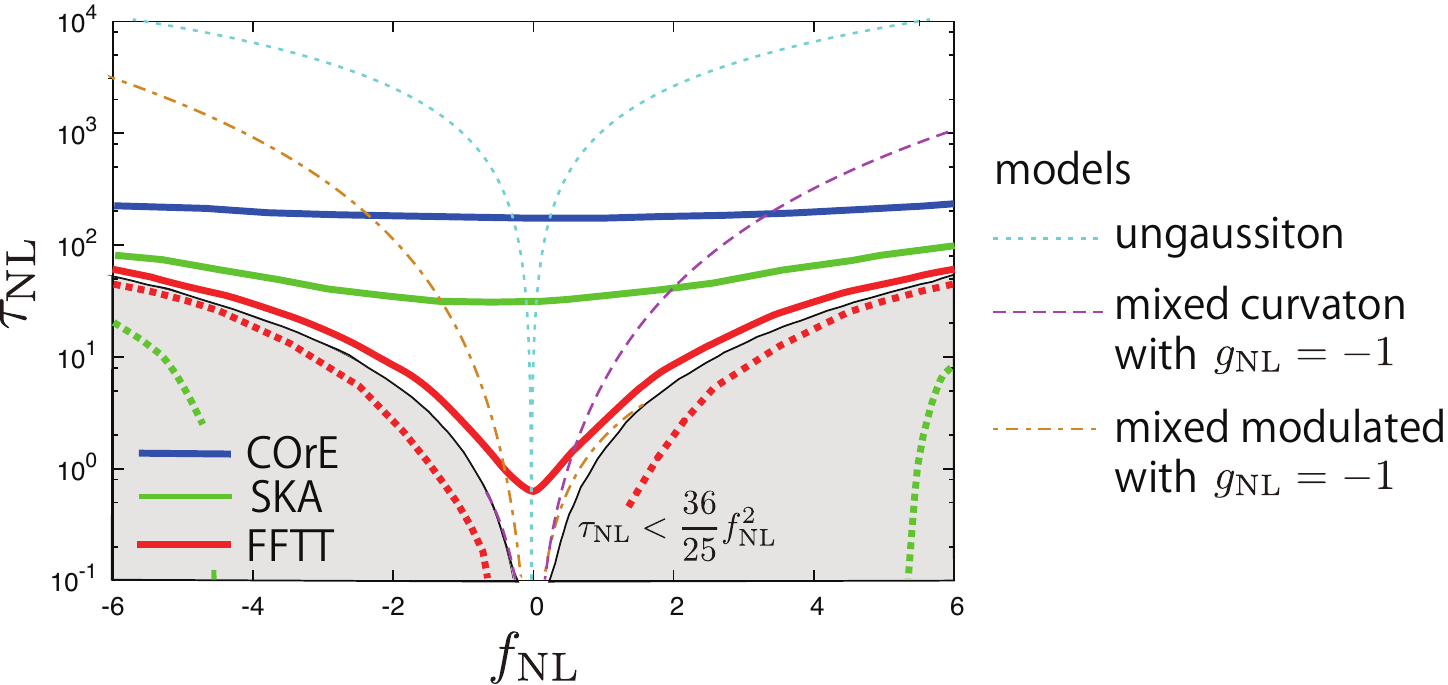}
\caption{\label{fig:consistency} 
Regions where the consistency relation $\tau_{\rm NL} = (36/25) f_{\rm NL}^2$ can be excluded at 1$\sigma$.
}
\end{figure}

In Fig.~\ref{fig:consistency},  regions where the consistency relation $\tau_{\rm NL} = (36/25) f_{\rm NL}^2$ can be excluded 
at 1$\sigma$ are shown for COrE, SKA and FFTT alone analysis are shown. For the fiducial values of $f_{\rm NL}$ and $\tau_{\rm NL}$ 
above each solid line and below each dashed line, we can confirm that the consistency relation does not hold at 1$\sigma$. 
For reference, we also show predictions of some models (see \cite{Suyama:2010uj,Suyama:2013rol} for details of the models)
in the same figure.

\begin{table*}
  \begin{center}
    \begin{tabular}{lcccccccccc}
      \hline\hline
      &\multicolumn{3}{c}{Planck} &\quad\quad&\multicolumn{5}{c}{COrE} \\
      \hline
      band frequency [GHz] &100 & 147 & 217 &\quad\quad& 105 & 135 & 165 & 195 & 225 \\
      beam width $\Delta\theta$ [arcmin] & 9.9 & 7.2 & 4.9 &\quad\quad& 10.0 & 7.8 & 6.4 & 5.4 & 4.7 \\
      Temperature noise $\Delta_T$ [$\mu$K arcmin] & 31.3 & 20.1 & 28.5 &\quad\quad& 2.68 & 2.63 & 2.67 & 2.63 & 2.64 \\
      Polarization noise $\Delta_P$ [$\mu$K arcmin] & 44.2 & 33.3 & 49.4 &\quad\quad& 4.63 & 4.55 & 4.61 & 4.54 & 4.57 \\
      \hline\hline
    \end{tabular}
  \end{center}
  \caption{\label{tab:cmb} Specification of CMB surveys.}
\end{table*}
\begin{table}[htbp]
\centering
  \begin{tabular}{lcc}
    \hline\hline
    & SKA & FFTT \\
    \hline
    total effective area $A_{\rm tot}$ [m$^2$]& $10^5$ & $10^7$ \\
    bandwidth $\Delta \nu$ [MHz] & \multicolumn{2}{c}{$1$} \\
    beam width $\Delta \theta$ [arcmin] & \multicolumn{2}{c}{$9$} \\
    integration time $t$ [hour] & \multicolumn{2}{c}{$1000$} \\
    \hline\hline
  \end{tabular}
  \caption{\label{tab:21cm}The survey specifications for 21~cm observations}
\end{table}
\begin{table}[htbp]
\centering
  \begin{tabular}{lccc}
    \hline\hline
     dataset &  $\Delta f_{\rm NL}$ & $\Delta g_{\rm NL}/10^3$ & $\Delta \tau_{\rm NL}$\\
    \hline
    CMB alone &  \\
    \quad~Planck & 4.0 & 41 & 610 \\
    \quad~COrE & 2.0 & 18 & 160 \\
    \hline
    SKA & 1.4 & 2.3 & 28\\
    \quad+Planck & 1.3 & 2.3 & 28 \\
    \quad+COrE & 1.1 & 2.2 & 27 \\
    \hline
    FFTT & 0.51 & 0.79 & 0.59 \\
    \quad+Planck & 0.50 & 0.78 & 0.58 \\
    \quad+COrE & 0.48 & 0.75 & 0.58 \\
    \hline\hline
  \end{tabular}
  \caption{\label{tab:constraints} Constraints on $f_{\rm NL}$, $g_{\rm NL}$ and $\tau_{\rm NL}$ at 1$\sigma$ with other parameters being marginalized over.} 
\end{table}

In Table~\ref{tab:zdep}, we summarize the dependence of the constraint on $z_{\rm min}$
in the cases of SKA+Planck and COrE+FFTT.
As the reionization proceeds, minihalos start to be ionized by
background UV and host stars by molecular hydrogen cooling. 
Therefore there exists a theoretical uncertainty in the determination of $z_{\rm min}$.
However most of minihalos can
survive until the late stage of the reionization process~($z\sim 8$)~\cite{Iliev:2004mb,Hasegawa:2012uf}.
We in this paper adopt $z_{\rm min} = 6$ as a fiducial value, which could be allowed depending on reionization models.

\begin{table}[htbp]
\centering
      \begin{tabular}{llccc}
      \hline\hline
      & $z_{\rm min}$ & $\Delta f_{\rm NL}$ & $\Delta g_{\rm NL}/10^3$ & $\Delta \tau_{\rm NL}$\\
      \hline
	\multirow{4}{*}{Planck+SKA} 
	& 4 & 1.0 & 0.93 & 26\\
	& 6 & 1.3 &  2.2 & 28\\
	& 8 & 2.2 & 8.1 & 33\\
	& 10 & 3.8 & 14 & 58\\
	\hline
	\multirow{4}{*}{COrE+FFTT} 
	& 4 & 0.38 & 0.58 & 0.56\\
	& 6 & 0.48 & 0.75 & 0.58\\
	& 8 & 0.63 & 0.98 & 0.61\\
	& 10 & 0.83 & 1.2 & 0.67\\
      \hline\hline
    \end{tabular}
  \caption{\label{tab:zdep} $z_{\rm min}$-dependence of the constraint on $f_{\rm NL}$, $g_{\rm NL}$ and $\tau_{\rm NL}$.}
\end{table}

\bigskip
{\it Discussion}.---\quad
As shown in Table~\ref{tab:constraints}, future observations of 21~cm fluctuations can probe the non-linearity parameters, 
particularly those for trispectrum as $\tau_{\rm NL} \sim 30$ and $g_{\rm NL} \sim 2 \times 10^3$ for SKA and 
$\tau_{\rm NL} \sim 0.6$ and $g_{\rm NL} \sim 8 \times 10^2$ for FFTT. 
As mentioned in the introduction, some inflationary models 
can generate a large value of $g_{\rm NL}$ while keeping $f_{\rm NL}$  small \cite{Suyama:2010uj,Suyama:2013rol}.  
Such models can be excluded once we obtain the above level sensitivity in future observations. 
Moreover, even if $f_{\rm NL}$ and $g_{\rm NL}$ are severely constrained, we still cannot differentiate between single-field and multi-field models. 
Nevertheless, if we can also probe $\tau_{\rm NL}$ with a good sensitivity, we will be able to test  models by looking at the consistency relation: 
the equality $\tau_{\rm NL}=(36/25)f_{\rm NL}^2$ is satisfied for a single-field model, while the inequality $\tau_{\rm NL} > (36/25)f_{\rm NL}^2$
holds for multi-field models.  As  examples of multi-field models, we show the predictions of the $f_{\rm NL}$--$\tau_{\rm NL}$ relation for
ungaussiton model \cite{Boubekeur:2005fj,Suyama:2008nt}, mixed curvaton and inflaton model \cite{Langlois:2004nn,Lazarides:2004we,Moroi:2005kz,Moroi:2005np,Ichikawa:2008iq}
and mixed modulated reheating and inflaton model \cite{Suyama:2007bg, Ichikawa:2008ne}
in Fig.~\ref{fig:consistency}. As seen from the figure, SKA can differentiate mixed models when $ |f_{\rm NL}| > 2$. 
With the sensitivity of FFTT, even when $f_{\rm NL} < {\cal O}(1)$, we can differentiate multi-field nature of the model.

As demonstrated in this paper, future observations of 21~cm angular power spectrum from minihalo would be a powerful tool to 
probe primordial non-Gaussianity, especially the trispectrum. By using this probe, we can further elucidate the mechanism of the inflationary Universe.

\bigskip
This work is partially supported by JSPS KAKENHI Grant Number
15K05084 (TT), 17H01131 (TT), 15K17646 (HT), 17H01110 (HT), 15K17659 (SY)
JP15H02082 (TS), 18H04339 (TS), 18K03640 (TS),  
MEXT KAKENHI Grant Number 15H05888 (TT, SY), 18H04356 (SY).

\end{document}